\documentclass[a4paper]{jpsj-suppl}
\usepackage{txfonts} %Please comment out this line unless the txfonts package is available in your LaTeX system.
\usepackage{epstopdf}

\newcommand{\Lambpha}{{}^{\ \ 6}_{\Lambda\Lambda}\text{He}}
\newcommand{\LL}{\Lambda\Lambda}
\title{Lambda-Lambda interaction from two-particle intensity correlation
in relativistic heavy-ion collisions}

\author{Akira \textsc{Ohnishi}$^{1}$, Kenji \textsc{Morita}$^{1,2,3}$,
and Takenori \textsc{Furumoto}$^{4}$}

\inst{$^{1}$Yukawa Institute for Theoretical Physics,
Kyoto University, Kyoto 606-8502, Japan\\
$^{2}$Frankfurt Institute for Advanced Studies,
Ruth-Moufang-Str. 1, D-60438 Frankfurt am Main, Germany\\
$^{3}$Institute of Theoretical Physics, University of Wroclaw,
PL-50204 Wroc law, Poland\\
$^{4}$National Institute of Technology, Ichinoseki College,
Ichinoseki, Iwate 021-8511, Japan}

\email{ohnishi@yukawa.kyoto-u.ac.jp}

\recdate{\today}

\abst{
We investigate $\Lambda\Lambda$ interaction dependence
of the $\Lambda\Lambda$ intensity correlation
in high-energy heavy-ion collisions.
By analyzing the correlation data recently obtained by the STAR collaboration
based on theoretically proposed $\Lambda\Lambda$ interactions,
we give a constraint on the $\Lambda\Lambda$ scattering length,
$-1.25~\text{fm} < a_0 < 0$,
suggesting that $\Lambda\Lambda$ interaction is weakly attractive
and there is no loosely bound state.
In addition to the fermionic quantum statistics
and the $\Lambda\Lambda$ interaction,
effects of collective flow, feed-down from $\Sigma^0$,
and the residual source are also found to be important to understand the data.
We demonstrate that the correlation data favor
negative $\Lambda\Lambda$ scattering length
with the pair purity parameter $\lambda=(0.67)^2$ evaluated
by using experimental data on the $\Sigma^0/\Lambda$ ratio,
while the positive scattering length could be favored
when we regard $\lambda$ as a free fitting parameter.
}

\kword{$\Lambda\Lambda$ interaction, Two particle intensity correlation,
Relativistic heavy-ion collisions}

\begin{document}
%\vskip -10pt%
\rightline{Preprint: YITP-15-126}
\vskip -12pt
\maketitle

\section{Introduction}

$\Lambda\Lambda$ interaction is a key ingredient
%has been attracting attention
in several subjects of nuclear physics, high-energy particle physics
and astrophysics.
First, $\Lambda\Lambda$ interaction is closely related to the existence
of the dihyperon, referred to as the $H$ particle ($uuddss$).
The color magnetic interaction is strongly attractive in the $H$ channel,
and a deeply bound $H$ particle state was predicted
% in the bag model
by Jaffe in 1977~\cite{Jaffe}.
While many dedicated experiments have been performed,
no evidence of deeply bound $H$ particle is found.
In 2001, the double hypernucleus 
%${}^6_{\Lambda\Lambda}\text{He}$
$\Lambpha$
was observed in the Nagara event~\cite{Nagara,Nagara-Update}.
The binding energy from the $^4\text{He}+\Lambda+\Lambda$ threshold
is obtained as 
$B_{\Lambda\Lambda}(\Lambpha)=6.91~\text{MeV}$~\cite{Nagara-Update},
then a bound $H$ state with the binding energy
$B_H \equiv 2M_\Lambda-M_H> B_{\Lambda\Lambda}(\Lambpha)$
was ruled out.
Nevertheless, physicists never gave up hunting $H$.
In 2007, the KEK-E522 collaboration observed
a bump structure with $2\sigma$ significance
at 15 MeV above the $\Lambda\Lambda$ threshold
in the $\Lambda\Lambda$ invariant mass spectrum
from $^{12}\text{C}(K^-,K^+\Lambda\Lambda)$~\cite{KEK-E522}.
%which may signal the resonance $H$ pole.
%
In more recent experiments,
the Belle and ALICE collaborations have searched for $H$
% the peaks 
in the $\Lambda{p}\pi^-$ invariant mass spectra
in $e^+e^-\to\Upsilon^*\to{HX}$~\cite{Belle2013}
and $\text{Pb}+\text{Pb}\to{HX}$~\cite{ALICE2016}
reactions, but there is no signal observed.
Recently performed ab-initio calculations
%Ab initio calculations have been performed recently and 
show the existence of the bound $H$
in the SU(3) limit~\cite{LQCD,Haidenbauer:2011ah},
while the SU(3) breaking effects may be large to push it up
above the threshold.
%
% recent experiments
%Thus 
We still have a chance to find $H$
as a very loosely bound state below the $\Lambda\Lambda$ threshold
or as a resonance,
where $\Lambda\Lambda$ interaction is relevant.
Second, $\Lambda\Lambda$ interaction is important to constrain
baryon-baryon interaction models.
Since the one-pion exchange process is not allowed between $\Lambda\Lambda$,
the interaction range is relatively short.
Thus the low energy scattering parameters $(a_0,r_\text{eff})$
in model calculations scatter in a wide parameter range.
It is possible to fit the interaction strength to the bond energy 
$\Delta B_{\Lambda\Lambda}\equiv B_{\Lambda\Lambda}(\Lambpha)-2B_{\Lambda}({}^5_\Lambda\text{He})=0.67~\text{MeV}$~\cite{FG,HKMYY,HKYM2010,Nagara-Update},
but we cannot determine both $a_0$ and $r_\text{eff}$ from one number
$\Delta B_{\Lambda\Lambda}(\Lambpha)$.
Thirdly, $\Lambda\Lambda$ interaction is crucial
%one of the key interactions
in neutron star physics.
There are some young and cold neutron stars which cannot be explained
by the standard cooling mechanism of neutron stars,
the modified URCA process $NN \to NNe\bar{\nu}~(\text{or } e^+\nu)$.
In order to explain the variety of surface temperatures of neutron stars,
admixture of superfluid non-nucleonic fermions is favored~\cite{Tsuruta}.
Superfluidity of $\Lambda$ is a promising candidate,
and we need precise strength
of $\Lambda\Lambda$ interaction to deduce the gap.
$\Lambda\Lambda$ interaction is also important to solve the "hyperon puzzle".
Hyperons are expected to appear in the core of heavy neutron stars,
while the equations of state of neutron star matter with hyperons
are generally too soft to support $2 M_\odot$ neutron stars~\cite{MassiveNS}.
If the $\Lambda\Lambda$ interaction is repulsive enough at high densities,
it may be possible to support massive neutron stars.
Density dependence of $\Lambda\Lambda$ interaction
or the $\Lambda\Lambda{N}$ three-body force is accessible
by comparing the vacuum and in-medium interactions.

Contrary to its importance, 
experimental information on $\Lambda\Lambda$ interaction is limited.
%We are interested in 
The two particle intensity correlation 
of $\Lambda\Lambda$ pairs from heavy-ion collisions
at ultrarelativistic energies,
which can provide information on $\Lambda\Lambda$ 
interaction~\cite{CGreiner1989,AO2000,MFO2015}.
Hadron production yields are well described by thermal models
in high energy collisions,
and one can expect enough yield of exotic hadron production
including dihyperons~\cite{SchaffnerBielich:1999sy,ExHIC}.
The chemical freeze-out condition is already studied~\cite{Thermal}
for heavy-ion collisions
at the Relativistic Heavy-Ion Collider (RHIC)
at the Brookhaven National Laboratory
and the Large Hadron Collider (LHC) at CERN.
Two particle intensity correlation is given as the convolution
of the source function
and the squared relative wave function~\cite{CorrFormula},
then careful analysis of the correlation tells us
the nature of the pairwise interaction.
Recently, the STAR collaboration has measured the $\Lambda\Lambda$ correlation 
in Au+Au collisions at $\sqrt{s_{NN}}=200~\text{GeV}$
at RHIC~\cite{STAR}.
The obtained data significantly deviate from unity,
and from the correlation expected from the quantum statistical correlation
between fermions.
Thus we expect that the data contain information
on the $\Lambda\Lambda$ interaction.

In this proceedings,
we investigate the $\Lambda\Lambda$ interaction dependence
of the $\Lambda\Lambda$ correlation.
We calculate the correlation function
by using several $\Lambda\Lambda$ interactions
proposed so far~\cite{NHC,NSC,ESC08,Ehime,fss2,FG,HKMYY,HKYM2010}.
Given a model source function relevant to heavy-ion collisions,
we discuss the interaction dependence
of the correlation function
and find that the $\Lambda\Lambda$ interaction is weakly attractive,
$-1.25~\text{fm}<a_0<0$~\cite{MFO2015}.
We also discuss the feed-down effects on the $\Lambda\Lambda$ correlation.
Analyses by the STAR collaboration~\cite{STAR}
show the scattering length whose sign is different from ours.
We find that the favored sign of the scattering length
depends on assumptions on the pair purity
%the $\Sigma^0/\Lambda$ ratio 
adopted in the analyses.

%when we adopt the $\Sigma^0/\Lambda$ ratio
%measured in $p$+Be collisions at 28.5 GeV/$c$~\cite{Sullivan1987}
%or calculated by using the statistical model,
%we

\section{$\Lambda\Lambda$ correlation and interaction from heavy-ion collisions}

\begin{figure}[tbh]
\begin{center}
\includegraphics[width=15cm]{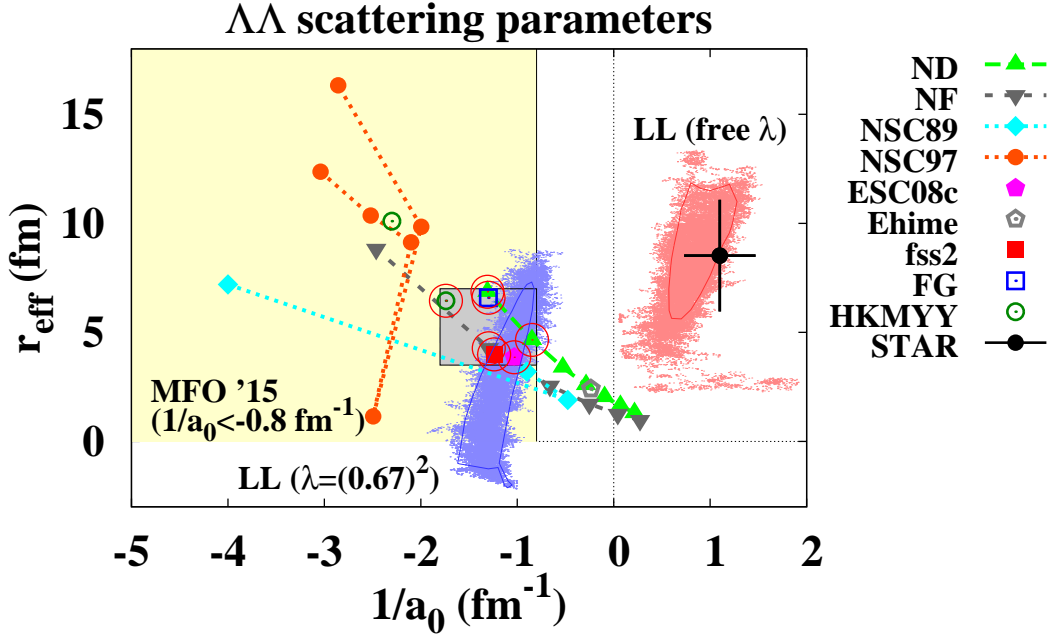}\\[2ex]
\end{center}
\caption{
Low-energy scattering parameters $(a_0,r_\text{eff})$ of $\LL$.
Symbols show $(1/a_0,r_\text{eff})$ from $\LL$
potentials~\cite{NHC,NSC,ESC08,Ehime,fss2,FG,HKMYY,HKYM2010},
and shaded areas show the region favored by the $\LL$ correlation data
in our work (MFO '15)~\cite{MFO2015}.
Dots show the Markov-chain Monte Carlo samples,
showing small $\chi^2/\text{DOF}$ region
in the LL model~\cite{LL81}
with $\lambda=(0.67)^2$~\cite{MFO2015} (blue dots)
and with $\lambda$ as a free parameter~\cite{STAR} (red dots).
Contours show $\chi^2/\text{DOF}=0.56$ ($\lambda=(0.67)^2$, red)
and $\chi^2/\text{DOF}=0.65$ (free $\lambda$, blue).
Filled black circle with $xy$ error bar shows
the analysis result by the STAR collaboration,
where $\lambda$ is regarded as a free parameter~\cite{STAR}.
}
\label{Fig:ar}
\end{figure}

There are many $\LL$ interaction proposed so far, such as
%and we compare the results from some of $\LL$ interaction models;
Nijmegen potentials (ND, NF, NSC89, NSC97, ESC08)~\cite{NHC,NSC,ESC08},
Ehime potential~\cite{Ehime},
Quark model potential (fss2)~\cite{fss2},
and Nijmegen-based potentials fitted to the Nagara data~\cite{FG,HKMYY,HKYM2010}.
Low energy scattering parameters $(a_0,r_\text{eff})$ of these potentials
are plotted in Fig.~\ref{Fig:ar}.
Most of these potentials do not predict the existence of the $\LL$ bound state.
Exceptions are the Nijmegen hard core models~\cite{NHC},
where the hard core radius is not given for $\LL$
and it is treated as a parameter.
%If we take 
For a small hard core radius, these models predict a bound state.
Recent $\LL$ potential models give similar low-energy scattering parameters,
$(a_0,r_\text{eff})=(-0.81~\text{fm}, 3.99~\text{fm})$ in fss2~\cite{fss2}
and
$(a_0,r_\text{eff})=(-0.97~\text{fm}, 3.86~\text{fm})$ in ESC08c~\cite{ESC08}.
$\LL$ potentials fitted to the Nagara event show larger effective ranges;
$(a_0,r_\text{eff})=(-0.77~\text{fm}, 6.59~\text{fm})$
in Filikhin-Gal (FG) potential~\cite{FG},
and
$(a_0,r_\text{eff})=(-0.58~\text{fm}, 6.45~\text{fm})$
in the potential given by Hiyama et al. (HKMYY)~\cite{HKMYY}.
When fitted to the updated value,
$\Delta{B}_{\LL}=0.67~\text{MeV}$~\cite{Nagara-Update},
the $\LL$ potential is predicted to be weaker,
$(a_0,r_\text{eff})=(-0.44~\text{fm}, 10.1~\text{fm})$~\cite{HKYM2010}.

The $\Lambda\Lambda$ correlation function from a chaotic source is given
as~\cite{CorrFormula}
\begin{align}
C(\bold{q})
=\frac{
\int d^4x_1 d^4x_2
S(x_1,\bold{p})
S(x_2,\bold{p})
\left| \psi^{(-)}(\bold{x}_{12},\bold{q}) \right|^2
}{
\int d^4x_1 d^4x_2
S(\bold{x}_1,\bold{p}+\bold{q})
S(\bold{x}_2,\bold{p}-\bold{q})
}
\ ,\label{Eq:Correlation}
\end{align}
where $S(x,\bold{p})$ is the source function of $\Lambda$ particle,
%$\bold{q}=(\bold{p}_1-\bold{p}_2)/2$ is the relative momentum,
$\bold{q}$ is the relative momentum,
$\bold{x}_{12}$ is the relative coordinate with time difference correction,
and $\psi^{(-)}$ is the "out" state wave function
with the asymptotic relative momentum $\bold{q}$.
%whose outgoing waves approaches those in $\exp(i\bold{q}\cdot\bold{r})$
%in the asymptotic region.
%where the relative momentum is $\bold{q}$ in the asymptotic region.
We assume here that only the $s$-wave is modified.
Then for the static and spherical source,
$S(x,\bold{p})\propto \exp(-\bold{x}^2/2R^2)\delta(t-t_0)$,
the correlation function is obtained as
\begin{align}
C_\text{sph}(\bold{q})\simeq
1 - \frac12 \exp(-4q^2R^2)+\frac12 \int_0^\infty dr S_{12}(\bold{r})
\left[
\left|\chi_0(\bold{r})\right|^2
-\left|j_0(qr)\right|^2
\right]
\ ,\label{Eq:corr1}
\end{align}
where $\chi_0$ is the relative wave function in the s-wave, 
$j_0$ is the spherical Bessel function,
and $S_{12}(\bold{r})=r^2\exp(-r^2/4R^2)/2\sqrt{\pi}R^3$
is the normalized source function in the relative coordinate.
The second term in Eq.~\eqref{Eq:corr1} is
the Hanbury Brown, Twiss (HBT) or Goldhabar, Goldhaber, Lee, Pais (GGLP) term,
which shows the suppression of the correlation
due to the anti-symmetrization of the wave function for fermions.
The third term shows the interaction effects;
when the wave function is enhanced due to the attraction,
the correlation is enhanced accordingly.

Let us examine the interaction dependence of the correlation function.
We show here the results of an analytic model developed
by Lednicky and Lyuboshits (LL)~\cite{LL81},
\begin{align}
C_\text{LL}(Q)=&1 - \frac12 e^{-R^2Q^2}
% + C_\text{int}(Q)
%\ ,\nonumber\\
%C_\text{int}(Q)=&
+\frac{|f(q)|^2}{4R^2}
%F_3\left(\frac{r_\text{eff}}{R}\right)
F_3(r_\text{eff}/R)
+\frac{\text{Re}f(q)}{\sqrt{\pi}R} F_1(QR)
-\frac{\text{Im}f(q)}{2R} F_2(QR)
%\ ,\nonumber\\
%f(q)=&(-1/a_0+r_\text{eff}q^2/2-iq)^{-1}
\ ,%\nonumber\\
\label{Eq:LL}
\end{align}
where $Q=2q$,
$f(q)=(-1/a_0+r_\text{eff}q^2/2-iq)^{-1}$ is the scattering amplitude,
$F_1(z)=\int_0^z dx e^{x^2-z^2}/z$,
$F_2(z)=(1-e^{-z^2})/z$,
and $F_3(x)=(1-x/2\sqrt{\pi})$.
It should be noted that we take the "nuclear physics" convention
for the scattering length, 
$q\cot\delta=-1/a_0 + r_\text{eff}q^2/2+\mathcal{O}(q^4)$,
which leads to $\delta \simeq -a_0 q$ at low energy.
This LL formula with $F_3=1$ is derived by using the asymptotic wave function
in the $s$-wave
$\chi_0\simeq e^{-i\delta}\sin(qr+\delta)/qr$
and a static spherical Gaussian source.
The $F_3$ term shows the effective range correction.
We find that the function $F_1$ is well approximated in the form
$F_1(z)=(1+c_1x^2+c_2x^4+c_3x^6)/(1+(c_1+2/3)x^2+c_4x^4+c_5x^6+c_3x^8)$
with $(c_1,\cdots,c_5)=(0.123,0.0376,0.0107,0.304,0.0617)$
in the $z$ range of interest, $0<z<20$.

\begin{figure}[tbh]
\begin{center}
\includegraphics[width=12cm]{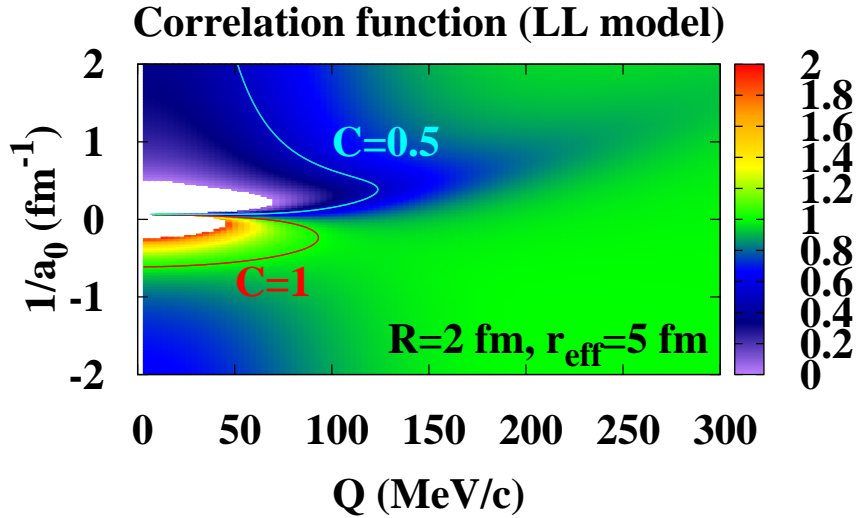}%
\end{center}
\caption{Correlation function $C_{\LL}$ as a function of $Q$ and $1/a_0$
in the LL model~\cite{LL81}.
We show the results for $R=2~\text{fm}$ and $r_\text{eff}=5~\text{fm}$
as an example.
}
\label{Fig:3DCLL}
\end{figure}

In Fig.~\ref{Fig:3DCLL}, we show the correlation function
for $R=2~\text{fm}$ and $r_\text{eff}=5~\text{fm}$.
When $|a_0|$ is small, the correlation function is approximately described
by the HBT term, and converges to 0.5 at $Q\to 0$.
In the negative $a_0$ case (attractive potential without loosely bound states),
the correlation function is enhanced especially at small $Q$,
because of the enhanced wave function by the attraction.
We note that when the scattering length is positive,
the correlation is generally suppressed;
%the wave function in the $s$-wave is given as
%$\chi_0\simeq e^{-i\delta}\sin(qr+\delta)/qr$ and $\delta\simeq -qa_0 <0$
%for $a_0>0$,
%then the squared wave function is small at around $r\simeq a_0$.
Positive $a_0$ means that there is a shallow bound state
or the interaction is repulsive,
then the squared wave function is suppressed by the node or by the repulsion.
Thus the correlation function is sensitive to the $\Lambda\Lambda$ interaction,
as long as other effects do not wash out the above trend.

We calculate the correlation function using the original formula
Eq.~\eqref{Eq:Correlation} rather than using the LL formula.
With a static spherical source,
we have searched for the optimal source size $R$
for each potential model.
The comparison with the STAR data tells that
the chi-square is not very good ($\chi^2_\text{min}/\text{DOF} \sim 2$)
and the optimal source size is small ($R = (1-1.5)~\text{fm}$).
We can take account of the collective flow effects 
by modifying the source function in Eq.~\eqref{Eq:Correlation}.
%
%We have taken account of the collective flows,
%and feed-down effects from $\Sigma^0\to\Lambda\gamma$,
%and the residual correlation
%for more serious estimate~\cite{MFO2015}.
%With the static spherical source with other 
%First, we consider the collective flows.
The Bjorken expansion is assumed for the longitudinal flow,
and the transverse flow strength is fixed by fitting the transverse 
momentum spectrum of $\Lambda$.
The chi-square becomes better with flow effects
($\chi^2_\text{min}/\text{DOF} \sim 1.5$),
while the optimal source size is still small ($R = (0.7-1.1)~\text{fm}$).
$\LL$ potentials with 
$-1.8~\text{fm}^{-1} < 1/a_0 < 0.8~\text{fm}^{-1}$
and $3.5 \text{fm} < r_\text{eff} < 7 \text{fm}$
are found to give reasonable fit to the data ($\chi^2/\text{DOF}<5$)
when we take the collective flow effects into account.
This region of scattering parameters are marked by the grey area
in Fig.~\ref{Fig:ar}.

\begin{figure}[tbh]
\begin{center}
\includegraphics[width=8cm]{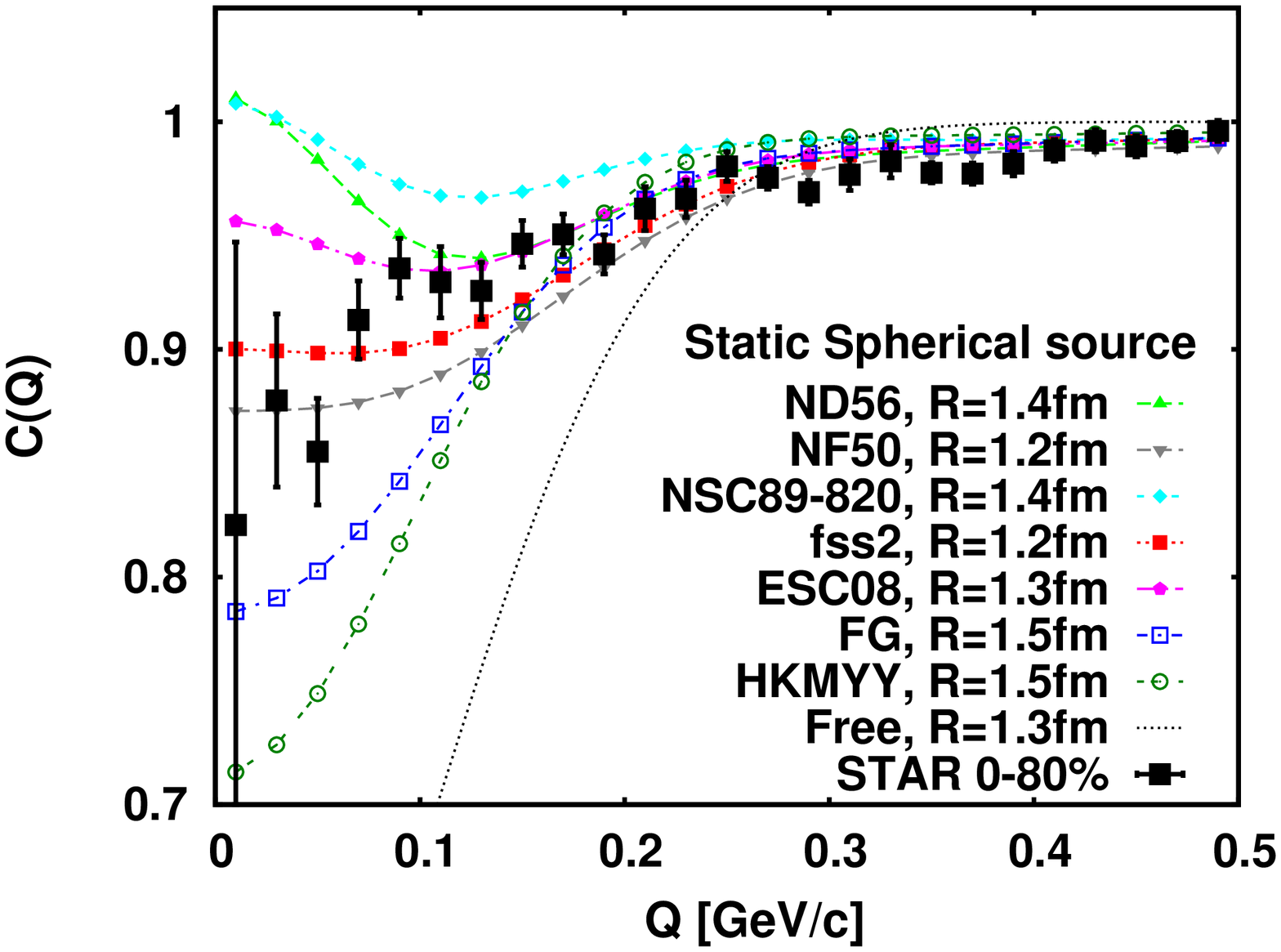}%
\includegraphics[width=8cm]{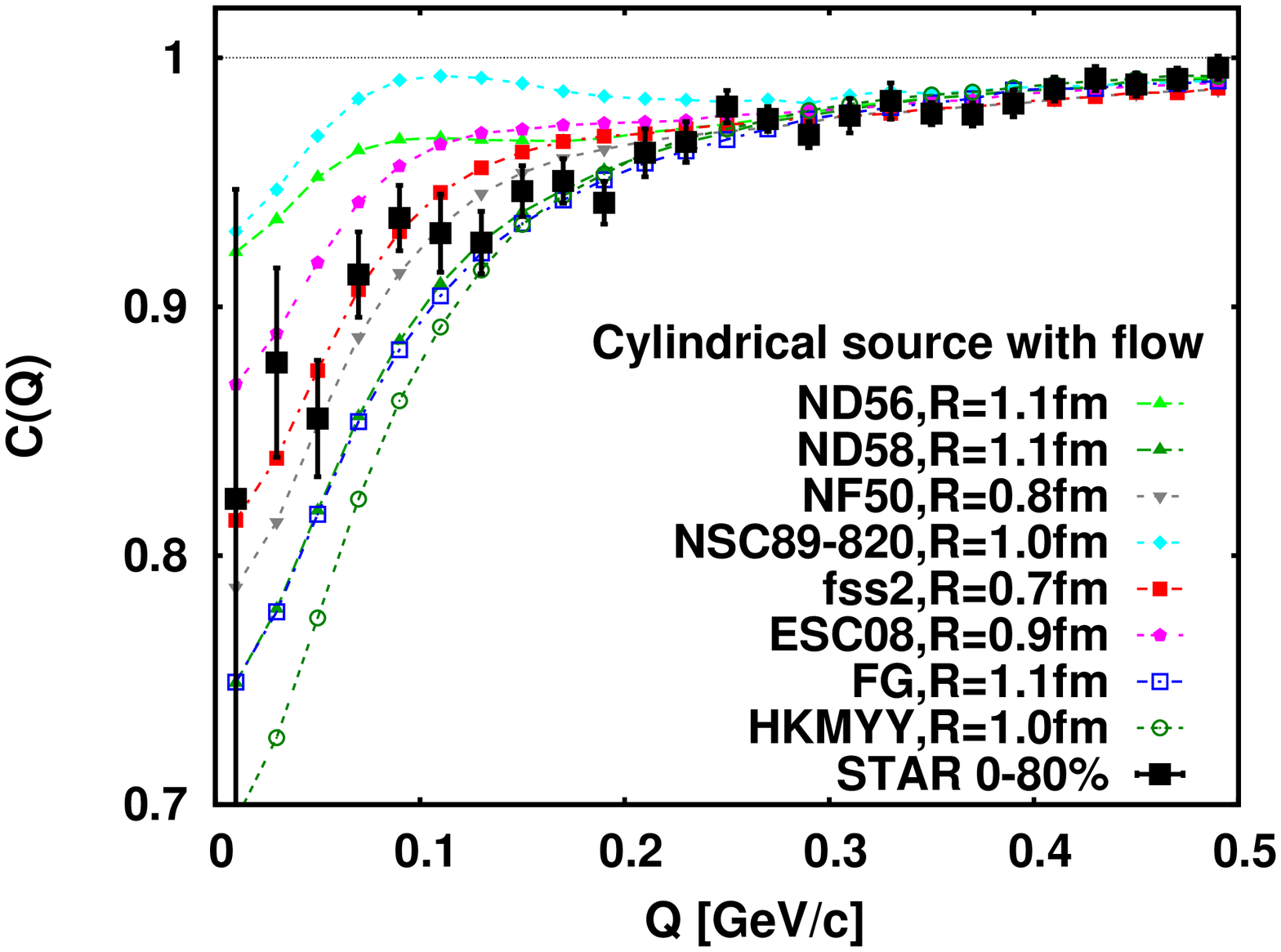}\\
\includegraphics[width=8cm]{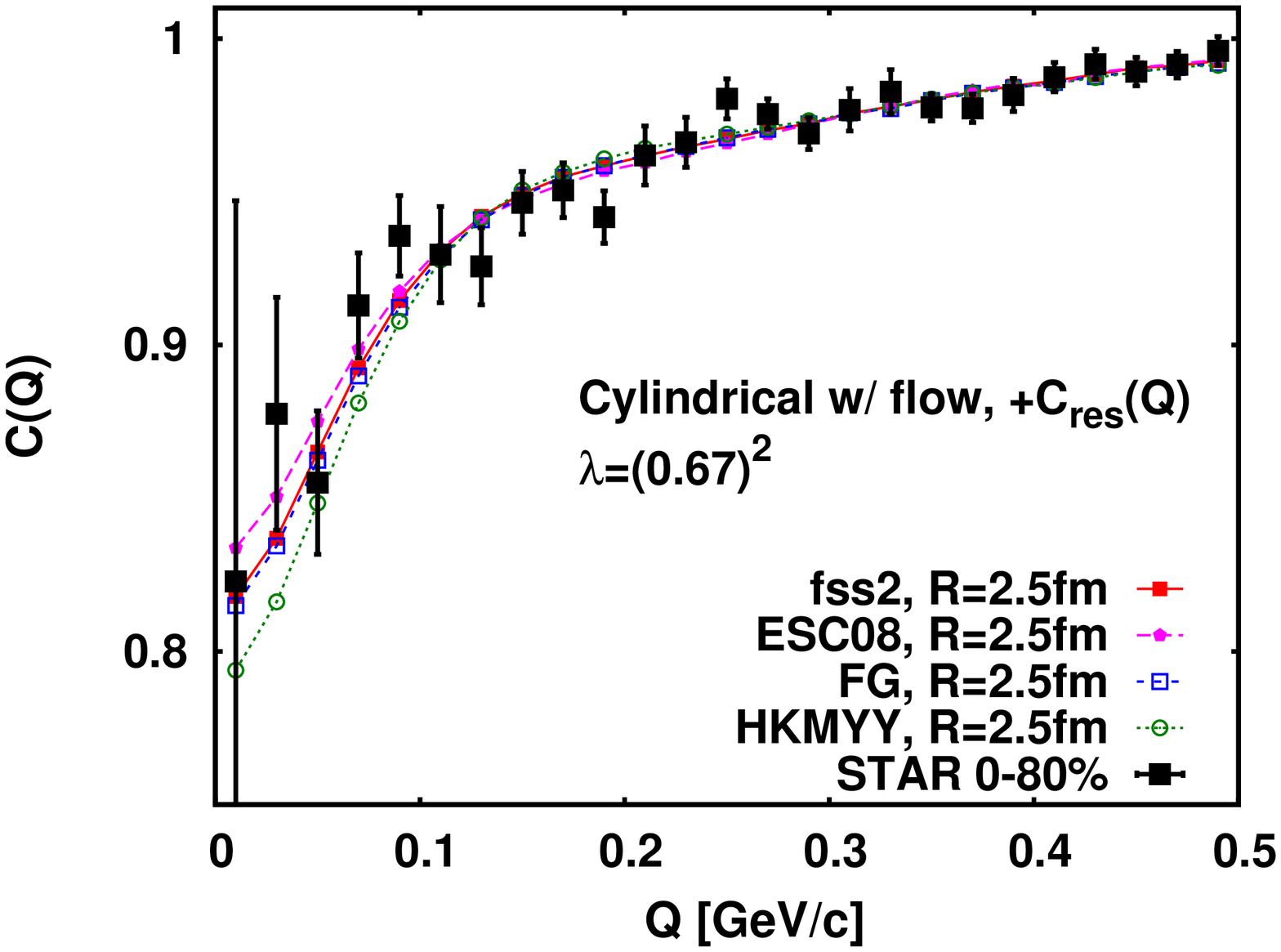}%
\end{center}
\caption{Correlation function obtained with several $\LL$
potentials~\cite{NHC,NSC,ESC08,Ehime,fss2,FG,HKMYY,HKYM2010}
in comparison with data~\cite{STAR}.
Upper left (right) panel shows the result
from the static and spherical source (cylindrical source with flow),
and the lower panel shows the result
with the cylindrical source including flow,
residual source and feed-down effects.
}
\label{Fig:CLL}
\end{figure}

Next, we consider the feed-down effects and "residual" source effects.
The discussion so far applies to the case where $\Lambda$ particles are
directly emitted from the hot matter.
However, decay from long-lived resonances,
$\Xi$, $\Omega$, and $\Sigma^0$, produces $\Lambda$ particles
of which correlations can be neglected
due to large pair separation and finite bin width. 
%In the data, we may also have $\Lambda$
%from the decay of $\Xi$, $\Omega$ and $\Sigma^0$ baryons,
%and these $\Lambda$s are not correlated.
Feed-down from short-lived hyperon resonances can be taken into account
by modifying the source size.
Since the lifetimes of $\Xi$ and $\Omega$ are not very small
($c\tau(\Xi)=8.71~\text{cm}$, $c\tau(\Omega)=2.46~\text{cm}$),
it is possible to reject $\Lambda$ from the weak decay
by using the distance of closest approach to the primary vertex~\cite{STAR}.
By contrast, $\Sigma^0$ decays electromagnetically,
and we cannot reject $\Lambda$ from $\Sigma^0$ decay,
which modifies the pair purity probability
such that the observed correlation function becomes 
%For a pair purity probability $\lambda$,
%the observed correlation function with the feed-down effects becomes
\begin{align}
C_\text{corr}(Q)=1+\lambda(C_\text{bare}(Q)-1)\ ,
\label{Eq:feed-down}
\end{align}
where $C_\text{bare}$ is calculated by using Eq.~\eqref{Eq:Correlation}.
We have adopted the pair purity $\lambda=(0.67)^2=((1-0.278-0.15)/(1-0.15))^2$
based on the observed ratio
$\Sigma^0/\Lambda_\text{tot}=0.278$~\cite{Sullivan1987}
and $\Xi/\Lambda_\text{tot}=0.15$~\cite{RHIC-Xi},
where $\Lambda_\text{tot}$ represents $\Lambda$ yield including
decay contributions.
While the above $\Sigma^0/\Lambda$ ratio is measured
in a different reaction, it is close to the statistical model estimate
and small modification of $\lambda$ does not change our conclusion.
Another important effect is the "residual" source.
In the STAR data, we find that the $\LL$ correlation function is suppressed
significantly even at high relative momentum region, $Q \sim 0.4~\text{GeV}$.
We do not know its origin, and assume that its effect is represented
by a Gaussian, $C_\text{res}=a_\text{res}e^{-r_\text{res}^2Q^2}$.
When we include the feed-down and residual source effects
in addition to the flow effects,
the $\LL$ correlation data is well explained.
In Fig.~\ref{Fig:CLL}, we show the comparison of the calculated
$\LL$ correlation function with fss2, ESC08, FG and HKMYY potentials
in comparison with data.
We find $\chi^2/\text{DOF}\simeq 1$
almost independent of the transverse size parameter in the range $R>0.5~\text{fm}$
for potentials with $1/a_0 \leq -0.8~\text{fm}^{-1}$
($-1.25~\text{fm}<a_0<0$),
marked by the yellow area in Fig.~\ref{Fig:ar}.

\section{$\LL$ scattering length; negative or positive ?}

As described in the previous section,
we have concluded that the $\LL$ scattering length is negative
and in the range $1/a_0<-0.8~\text{fm}^{-1}$.
By contrast, the STAR collaboration obtained a different result;
$a_0=1.10\pm 0.37^{+0.08}_{-0.68}~\text{fm}$.
This positive scattering length suggests that
there is a bound state of $\LL$ or the $\LL$ interaction is repulsive.
Neither of these conclusions are not immediately acceptable,
then we now discuss the reason of the difference.

One of the differences between our analyses~\cite{MFO2015}
and the STAR collaboration analyses~\cite{STAR}
is the assumption on the pair purity probability $\lambda$.
We have evaluated its value based on the measurement
of $\Sigma^0$ and $\Xi^0$,
while the STAR collaboration takes $\lambda$ as a free parameter.
Another difference is the correlation function formula;
we have used the original formula Eq.~\eqref{Eq:Correlation},
while the STAR collaboration has used the LL formula, Eq.~\eqref{Eq:LL}.
In order to pin down which is essential,
we re-analyze the data by using the LL formula
with different assumptions on $\lambda$.

\begin{figure}[tbh]
\begin{center}
\includegraphics[width=8cm]{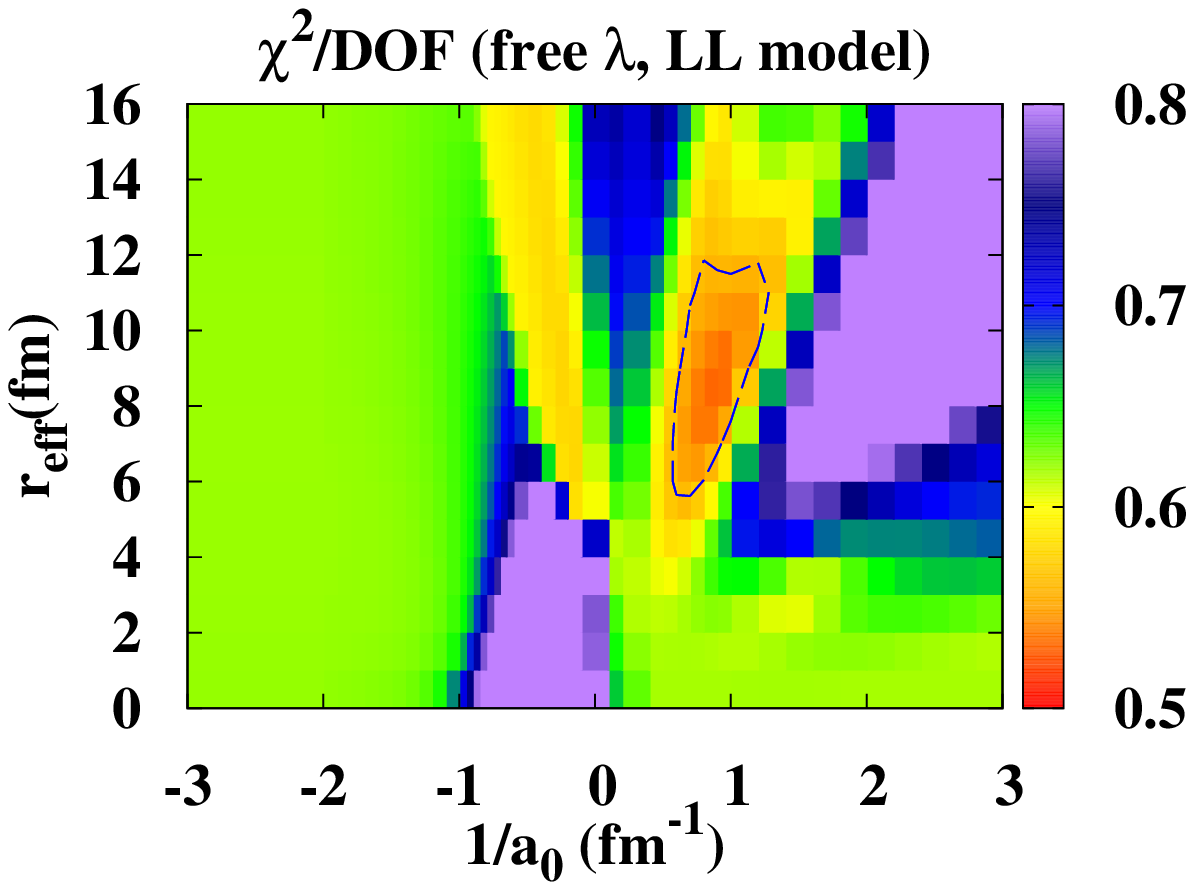}%
\includegraphics[width=8cm]{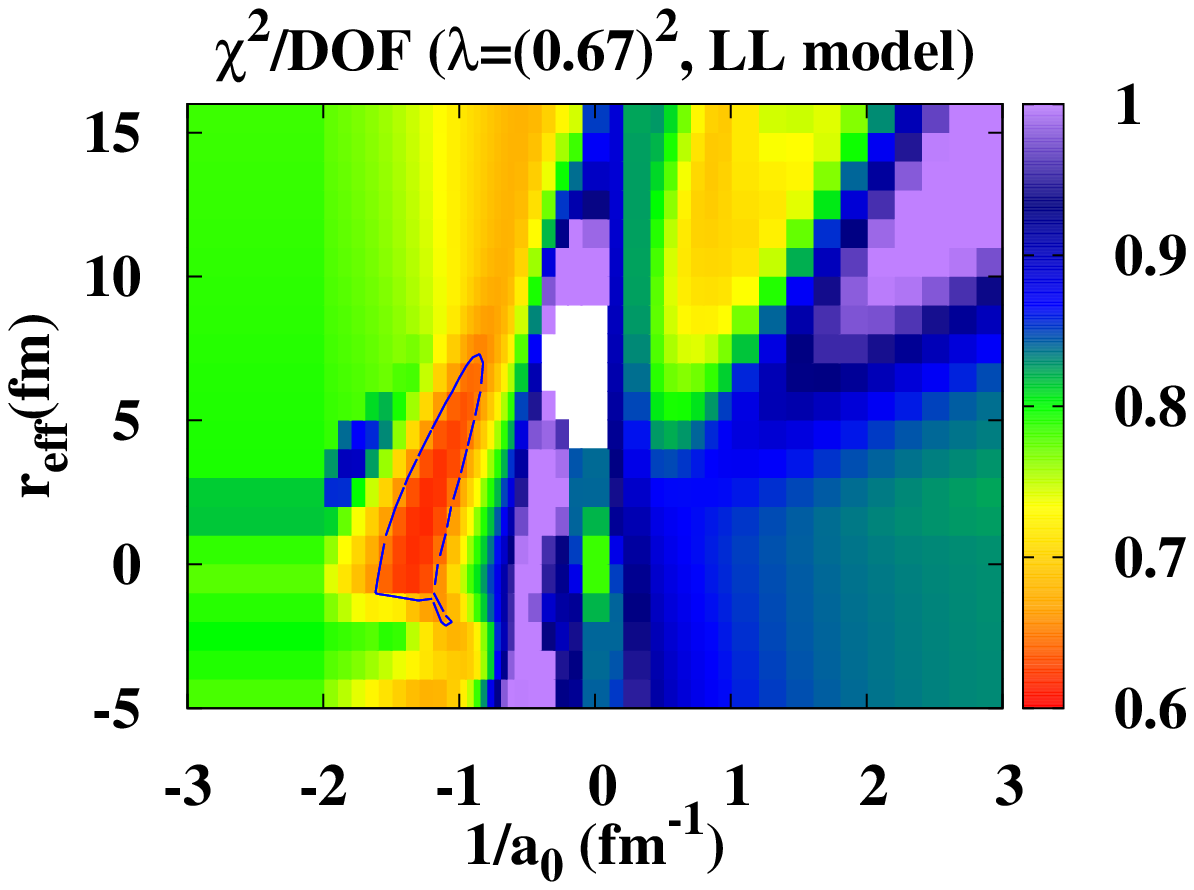}
\end{center}
\caption{
Contour plot of $\chi^2/\text{DOF}$ as a function 
of $1/a_0$ and $r_\text{eff}$ in the LL model.
In the left panel, we show the results
with $\lambda$ regarded as a free parameter.
In the right panel, results with $\lambda=(0.67)^2$ are shown.
}
\label{Fig:complambda}
\end{figure}

In Fig.~\ref{Fig:complambda},
we show the $\chi^2/\text{DOF}$ as a function of $1/a_0$ and $r_\text{eff}$
with $\lambda$ as a free parameter (left)
and
with a fixed $\lambda$ value, $\lambda=(0.67)^2$ (right).
We take the residual source effects into account.
When the $\lambda$ value is chosen to minimize the $\chi^2$,
we find $\lambda\simeq 0.18$ and the optimal $a_0$ value becomes negative,
as given by the STAR collaboration;
Quantum statistics and the pair purity give
$C(Q\to0)=1-\lambda/2 \sim 0.91$ at $\lambda=0.18$
from Eq.~\eqref{Eq:feed-down},
while the data show $C(Q\to0) \simeq 0.82$.
Thus with $\lambda=0.18$, we need to suppress $C(Q)$ at small $Q$
and positive $a_0$ is required.
%%%%%%%%%%%
By contrast, for a fixed $\lambda=(0.67)^2$,
the optimal $a_0$ value is found in the negative region,
as we have concluded in the previous section and in Ref.~\cite{MFO2015}.
We have also performed a Markov-chain Monte Carlo (MCMC) simulation
with the $\chi^2$ as the action.
In Fig.~\ref{Fig:ar}, dots show the sampled points.
MCMC results are consistent with the $\chi^2$ fit results;
The red contour shows $\chi^2/\text{DOF}=0.56$
with $\lambda$ as a free parameter,
and MCMC samples (red dots) scatter around this area.
The blue contour shows $\chi^2/\text{DOF}=0.65$
with $\lambda=(0.67)^2$,
where most of the MCMC samples (blue dots) are inside the contour.

The above analyses demonstrate the importance of understanding
the feed-down contribution.
Once we can fix the $\Sigma^0/\Lambda$ ratio experimentally,
it becomes possible to determine the sign of $a_0$ more definitely.
We would like to claim that
our assumption on the $\lambda$ value seems to be reasonable;
it is based on the direct measurement,
consistent with the statistical model estimate,
and the resultant negative scattering length is consistent
with the Nagara event as well as recently proposed
$\Lambda\Lambda$ potentials, fss2 and ESC08.

%\appendix
%\section{}

\section{Summary}

We have analyzed the $\LL$ intensity correlation in high-energy heavy-ion 
collisions, which has sensitivity to the $\LL$ interaction~\cite{MFO2015}.
Our analysis of the recently obtained STAR data implies that
the favored $\LL$ interaction has negative scattering length
$1/a_0 < -0.8~\text{fm}^{-1}$ ($-1.25~\text{fm}<a_0<0$),
where we take a convention $\delta = -a_0 q$ at low energy.
Our result is consistent with the Nagara event.
The observed $\LL$ correlation can be understood by taking account of
the anti-symmetrization of the wave function (HBT effects),
$\LL$ interaction,
collective flow effects,
feed-down effects,
and the residual source contribution.

The difference between the STAR collaboration analyses~\cite{STAR}
and ours~\cite{MFO2015}
lies in the assumption on the pair purity parameter $\lambda$,
which is known also as the chaoticity or the intercept parameter.
We have assumed that $\Sigma^0/\Lambda \sim 0.67$
and that one can reject $\Lambda$ particles from the weak decay of
$\Xi$ and $\Omega$ hyperons in experiment.
Under this assumption,
$\LL$ potentials with negative scattering length are found to be favored
by the correlation data
in the model proposed by Lednicky and Lyuboshits~\cite{LL81},
which has been used in the STAR collaboration analyses.
When $\lambda$ is regarded as a free parameter,
$\chi^2$ becomes smaller but a positive scattering length is favored.
Experimental confirmation of $\Sigma^0$ yield is important.

The present analysis implies that we can investigate
hadron-hadron interaction from two particle intensity correlation.
For instance, $\Omega{N}$ correlation has been studied in \cite{Hyp2015-Morita},
%Preliminary results on $\Omega^-N$ correlation at RHIC
%has been shown in Ref.~\cite{MOH2015},
and correlations of various hadron-hadron pairs including $\Omega^-N$
may be available at RHIC.
As for the $\LL$ interaction,
comparison with data obtained 
at the B-factory~\cite{Belle2013} and LHC~\cite{ALICE2016}
as well as data to be obtained at J-PARC~\cite{J-PARC_E42} should 
be helpful to constrain $\LL$ interaction more precisely.
Understanding the origin of the "residual" source is a theoretical challenge;
it may come from the color flux tube,
or it may be the signal of $p$-wave repulsion of $\LL$.

\section*{Acknowlegement}
%The authors would like to thank N. Shah and H. Z. Huang
%for providing them the STAR data.
%They also would like to acknowledge T. Rijken, S. Shinmura, Y. Yamamoto,
%E. Hiyama, A. Gal, Y. Akaishi and B. M$B!/(Buller for
%helpful discussions. 
%K.M. would like to thank K. Redlich
%for discussion on $\Sigma^0$ decay and hyperon yields in thermal
%models. He also acknowledges other members of Institute
%of Theoretical Physics in University of Wroclaw
%for discussion.
Numerical computations were carried out on SR16000 at YITP in Kyoto university.
This work is supported in part 
by the Grants-in-Aid for Scientific Research from JSPS
(Nos. 15K05079, %((C) AO, Y.Nara)
and   15H03663), %((B) PI:A.Nakamura) ), 
by the Grants-in-Aid for Scientific Research on Innovative Areas from MEXT
(Nos. 24105001, 24105008),
by the Yukawa International Program for Quark-Hadron Sciences,
and 
by HIC for FAIR and by the Polish Science Foundation (NCN),
under Maestro grant 2013/10/A/ST2/00106.

\end{document}